# Ultrafast Magnetization Reversal by Picosecond Electrical Pulses

Yang Yang[1],*, R. B. Wilson[4],*, Jon Gorchon[2,3],*, Charles-Henri Lambert[2],

Sayeef Salahuddin[2,3] and Jeffrey Bokor[2,3].

1) Department of Materials Science and Engineering, University of California, Berkeley, CA 94720, USA
2) Department of Electrical Engineering and Computer Sciences, University of California, Berkeley, CA 94720, USA
3) Lawrence Berkeley National Laboratory, 1 Cyclotron Road, Berkeley, CA 94720, USA
4) Department of Mechanical Engineering and Materials Science and Engineering Program, University of California, Riverside, CA 92521, USA
*denotes equal contributions
correspondence should be addressed to y-yang@berkeley.edu, rwilson@engr.ucr.edu or jbokor@berkeley.edu.

The field of spintronics involves the study of both spin and charge transport in solid state devices with a view toward increasing their functionality and efficiency [1]. Alternatively, the field of ultrafast magnetism focuses on the use of femtosecond laser pulses to excite electrons in magnetic materials, which allows the magnetic order to be dramatically changed on unprecedented sub-picosecond time-scales [2]. Here, we unite these two distinct research activities by using picosecond electrical pulses to rapidly excite electrons in a magnetic metal. We are able to deterministically and repetitively reverse the magnetization of a GdFeCo film with sub-10 picosecond electrical pulses. The magnetization reverses in ~10ps, which is more than an order of magnitude faster than any other electrically controlled magnetic switching[3-6]. We attribute the deterministic switching of the magnetization to ultrafast excitation of the electrons, a fundamentally different mechanism from other current driven switching mechanisms such as spin-transfer-torque (STT) or spin-orbit-torque (SOT). The energy density required for switching is measured and the process is found to be efficient, projecting to only 4 fJ needed to switch a (20 nm)$^3$ cell, which is comparable to other state-of-the-art STT-MRAM memory devices. This discovery will launch a new field of research into picosecond spintronic phenomena and devices.

**Main Text**

Spintronic devices are particularly promising candidates for future low-energy electronics taking advantage of the non-volatility of nanoscale magnets. For example, magnetic random access memory (MRAM) is now emerging as a universal integrated on-chip memory[7]. Spintronic logic devices are also being actively investigated due to their potential for low-power computing[8]. A significant obstacle that impedes the wide spread adoption of spintronic devices is their speed. The speed of spintronic devices, such as STT or SOT memories, is limited by the precessional period of the magnetization of the magnetic medium. Precessional periods of magnetic materials are typically on the order of hundreds of picoseconds [3]. For comparison, MOSFET transistors can have switching delays as short as 5 ps [9]. In order for spintronic technologies to challenge charge based devices in information technologies, the speed of operation must be dramatically improved.

Research over the past two decades in the field of ultrafast magnetism demonstrates that precessional speed limits for manipulating magnetic order can be broken if the electrons are excited on time-scales faster than the electron-phonon relaxation time, i.e. excited on sub-ps time-scales [2]. For example, the magnetization of a ferromagnetic thin film can be quenched within 300 fs upon 60 fs laser irradiation [10]. Furthermore, multiple studies have demonstrated the ability of single 100 fs laser pulses to deterministically and repetitively switch the magnetization of the ferrimagnetic metal GdFeCo on sub-picosecond time-scales [11-14], a phenomenon known as all optical switching (AOS).

The driving force behind all optical switching of ferrimagnetic metals is ultrafast heating of the electronic system[12-14]. Ferrimagnets such as GdFeCo contain two distinct spin sublattices aligned antiparallel to each other with uncompensated magnetic moments. Picosecond switching is understood as due to the distinct dynamics of the Gd and the FeCo sublattices [12]. First, the magnetic sublattices undergo demagnetization at different rates after laser excitation[12]. Then, angular momentum exchange between sublattices enables a transient alignment of the two sublattices, followed by picosecond magnetization reversal [12].

Here, we take advantage of the physics responsible for AOS in GdFeCo to demonstrate a new regime of purely electrical ultrafast spintronics. Instead of optically exciting electrons, we use picosecond charge current pulses to excite the electrons of a GdFeCo metal film. We observe repeatable ultrafast magnetization reversal in the GdFeCo film with single sub-10 ps electrical pulse excitation.

To generate picosecond electrical pulses, we fabricated picosecond photoconductive switches [15], in a gold coplanar stripline (CPS) geometry, on a "low temperature (LT) grown" GaAs substrate [16]. Figure 1.a shows a schematic of the CPS device. The CPS is tapered from 50 μm to 5 μm, contacting on top of both sides of a patterned GdFeCo film, leaving a 4 μm by 5 μm uncovered GdFeCo section. By illuminating the DC biased photoconductive switch with 60 fs laser pulses at 810 nm wavelength, we are able to generate current pulses that have a duration of ~9 ps at full width half maximum (FWHM) and a current density up to ~$10^9$ A/cm² through the GdFeCo section (Figure 1.b). Additional information on the sample fabrication and current pulse characterization is contained in the methods and extended data sections.

In our experiments, we study the magnetic response of a Ta(5 nm) / $Gd_{30}Fe_{63}Co_7$(20 nm) / Pt(5 nm) multilayer. The film presented perpendicular magnetic anisotropy with a coercivity of 80 Oe at room temperature. The compensation temperature, i.e. the temperature where the net moment is minimized in ferrimagnets, was ~270 K (Extended Data Figure 1). Consistent with prior studies [12-14], as shown in the differential magneto optic Kerr effect (MOKE) images of Figure 2.a, upon irradiation by a sequence of single laser pulses, the magnetization of the GdFeCo film toggles after each pulse. We checked the AOS ability at different laser pulse durations and found the GdFeCo film switches with laser pulse durations between 60fs (FWHM) and 10 ps, consistent with the results reported in Ref. 14.

We examined the response of the GdFeCo film to electrical pulses. Figure 2.b shows the differential MOKE images of the device after a sequence of single 9 ps electrical pulses with current density ~$7*10^8$ A/cm$^2$ through the GdFeCo section. The magnetization of the GdFeCo section toggles after each electrical pulse, just as in the optical experiments. The switching behavior is driven by transient heating of the electrons, as described in prior studies of AOS in GdFeCo [13,14].

We performed time resolved MOKE measurements in order to temporally resolve the switching dynamics following the arrival of an electrical pulse. Figure 3.a shows the magnetic dynamics that result from electrical pulses of different amplitude. For electrical pulses with an absorbed energy density in the GdFeCo section less than 0.8 mJ/cm$^2$ (with reference to the surface area), the MOKE signal (primarily indicative of the FeCo sublattice magnetization, see the methods section) shows demagnetization within 20 ps, followed by a recovery to the initial magnetization state on longer time-scales. With increased current pulse amplitude, the FeCo demagnetization is larger. For the electrical pulses with an absorbed energy density greater than 1.3 mJ/cm$^2$, the magnetic moment of the FeCo sublattice reverses within ~10 ps of the electrical pulse arrival at the GdFeCo film. Following reversal, the FeCo magnetization recovers rapidly towards the opposite direction. It reaches 70% of saturation within just 30 ps. We attribute the non-monotonic evolution of the magnetization, e.g. the decreasing of magnetization at ~40 ps, to the arrival of several low amplitude electrical pulses that arise from reflections of the initial pulse from various electrical discontinuities in the CPS structure.

For comparison, we performed time resolved AOS experiments on the same material. We used optical pulses with a FWHM of either 1 ps or 6.4 ps. The latter pulse duration is equivalent to the duration of electrical heating from a 9 picosecond electrical pulse because Joule heating is proportional to square of the current. The temporal evolution of the FeCo magnetization following optical irradiation is shown in Fig 3.b. We also include in Fig. 3b the magnetization dynamics that results from a 9 picosecond electrical pulse. The 1 ps optical pulses switched the magnetization in ~4 ps. Both the 6.4 ps optical pulse and the 9 ps electrical pulse switch the magnetization in ~10 ps. While the time-scale for switching is comparable, significant differences exist between optical and electrical switching. After the magnetization crosses through zero, the recovery of the magnetization in the opposite direction takes hundreds of ps longer for optical vs. electrical switching. This indicates the system reaches a higher equilibrium temperature after optical pulse excitation, given that electrons, spins and lattice are almost certainly in thermal equilibrium after 50 ps [14]. We determined the absorbed critical fluences required for optical switching to be 1.2 mJ/cm² and 1.6 mJ/cm² for 1 ps and 6.4 ps durations, respectively. These fluence values for AOS are on the same order as our observations in previous work [14]. Shorter optical pulses require somewhat less energy to reverse the GdFeCo magnetization, consistent with prior studies [14]. We determined the critical energy deposited in the GdFeCo during the electrical switching experiment to be 1.3 mJ/cm² (details in the methods and extended data sections). Therefore, both our time-domain measurements of the magnetization and our estimates of the absorbed fluence indicate that electrical switching of GdFeCo requires less energy than optical switching.

    One possible explanation for the different energy requirements of electrical vs. optical switching is the thermal vs. nonthermal nature of electrical vs. optical heating. In a parallel work, we have demonstrated that thermal vs. non-thermal heating of ferromagnetic metals results in distinct magnetization dynamics [17]. The transfer of energy from electronic to spin degrees of freedom relies on scattering events, such as Elliot-Yafet scattering and electron-magnon scattering [18]. Total scattering rates will depend strongly on both the number of excited electrons, and the average energy of excited electrons [19]. Electrical heating results in a large population of excited electrons with average energies less than 10 meV, while optical heating initially excites a much smaller number of electrons with eV scale

energies [17]. The ability to induce magnetization reversal in GdFeCo by picosecond electrical heating demonstrates that exciting a nonthermal electron distribution is not necessary for magnetization reversal [20].

The ability to switch a magnetic metal such as GdFeCo with a short electrical pulse has significant potential technological impact. For experimental convenience, we used an optoelectronic switch to generate picosecond electrical pulses, however this is not necessary. It is currently possible to generate and deliver sub-10 ps current pulses on-chip in conventional CMOS electronics. For example, a 5 ps ring oscillator delay has been demonstrated with 45 nm CMOS technology [9]. Therefore, it should be possible to implement GdFeCo based ultrafast on-chip memory and logic devices. A memory device would also require electrical read out. The addition of an oxide tunnel junction to the GdFeCo stack would enable electrical read out of the magnetic state [21].

A non-volatile ultrafast embedded MRAM technology based on ultrafast electrical excitation of the GdFeCo electrons would not only allow low static power dissipation due to the non-volatility of GdFeCo, but also require low dynamic energy consumption. The energy density required to switch the magnetization in our device is 13 aJ/nm$^2$. For a cell-size of (20 nm)$^3$, which is typical for memory devices [22], switching should be possible with a current pulse with a peak current of 3 mA that delivers ~4 fJ of energy to the electrons. The energy required for switching remains low despite the high current density required because the electrical pulse duration is short. We conclude that picosecond electrical switching of GdFeCo can be as energy efficient as STT and SOT schemes [23-26], yet more than one order of magnitude faster. So far, the peak current requirements for switching that we observe represent significant technological challenges for practical implementation. Further device and structure optimization should allow for significant reductions in the peak current and energy required for electrical switching. For example, switching energy per unit area could be lowered by reducing the thickness of the GdFeCo stack from the 30 nm used here.

An important merit of MRAM over other memory devices is the nearly unlimited cycling endurance [7]. Electrical heating of GdFeCo shows strong potential for high endurance in our experiments. We observe no degradation of electrical or magnetic properties in our devices after more than 10 hours of pump-probe experiments, which were performed with a laser repetition rate of 252 kHz. Ten hours of experiments corresponds to more than $10^{10}$ cycles. Although the peak current density during switching is high (~$7*10^8$ A/cm$^2$), the current pulse duration is only 10 ps, and thus average current density is far lower. A lifetime in excess of $10^{10}$ cycles is many orders of magnitude higher than most resistive RAM, phase-change memory or conductive bridging RAM [27].

In summary, we demonstrate that picosecond electrical heating by charge current injection can reverse magnetic order efficiently, yet more than one order of magnitude faster than any other current induced method. Our discovery bridges the gap between the fields of spintronics and ultrafast magnetism, which we believe opens a new frontier of ultrafast spintronics science and related devices.

**Methods**

**Sample fabrication.** The LT GaAs substrate, (PAM-Xiamen), consists of a 1 μm thick GaAs layer grown at low temperature by molecular beam epitaxy on a GaAs substrate. Time domain thermal reflectance measurements show a carrier lifetime of around 1.4 ps. The device fabrication process consists of three photolithography, material deposition and lift-off steps. First, a 100 nm-thick MgO layer is deposited by RF sputtering and patterned by a standard lift-off process. The substrate is fully covered with MgO except for several 100 μm by 60 μm windows where the photoconductive switches are placed in order to electrically isolate the CPS from the substrate. As a second step, the magnetic layer (5 nm Ta/20 nm $Gd_{30}Fe_{63}Co_7$/5 nm Pt) is sputtered and patterned using the same methods as for the MgO layer. This defines a GdFeCo island 5 μm by 20 μm in size within the MgO window. As a third and final step, a coplanar stripline (CPS) consisting of 20 nm thick Ti and 250 nm thick Au is e-beam evaporated and patterned. The CPS design contains a tapered region on each side of the GdFeCo island that narrows down the width of the lines in order to increase the current density at the GdFeCo section. The narrow part of the CPS on each side overlaps with the edges of the GdFeCo island, allowing for electrical pulses to flow through it. More fabrication details can be found in the supplemental information of ref. 17.

**Electrical pulse generation and characterization.** When the CPS is DC biased, a 60fs (FWHM) laser pulse discharges the CPS by irradiating the photo-switch, hence generating an ultrafast electrical pulse propagating along the line. We use a THz probe (Protemics Teraspike) to characterize the temporal profile of the electric field on the CPS device. The probe consists of a 2 µm wide LT-GaAs photo-switch that is positioned on the end of a flexible Polyethylene terephthalate (PET) cantilever. We position the probe tip between the two lines of the CPS at the region of interest of the sample (Extended Data Figure 2). A probe laser beam illuminates the photo-switch on the tip. A linear delay stage controls the optical delay between the probe beam and the pump beam exciting the photo-switch on the sample. During probe beam illumination, the Protemics probe outputs a photocurrent proportional to the strength of the electric field between the lines of the CPS. By monitoring the average photocurrent from the Protemics probe as a function of the optical delay between the pump and probe lasers, we map the temporal profile of the electric field at that location on the CPS. We use the average photocurrent generated by the photoswitch on the CPS, as measured by a DC voltage source (Keithley 2410 Source-meter) to estimate the total charge contained in each electrical pulse. Using the measured temporal profile, we can determine the peak current amplitude of the electrical pulse.

**Single shot all-optical-switching of GdFeCo.** We used an external magnetic field to homogeneously polarize the magnetization in the out-of-plane direction. Then, we irradiated the GdFeCo with a single linearly-polarized 810 nm wavelength laser pulse. We used differential MOKE microscopy to image the magnetization direction before and after laser irradiation.

**Time Resolved MOKE measurement.** An amplified Ti:Sapphire laser with 252 kHz repetition rate is used for time resolved measurements. The 60 fs laser pulse, with 810 nm center-wavelength, is split into pump and probe beams. The probe is used to measure the magnetization of the sample via MOKE. The pump beam reflects off a retro-reflector on a linear delay stage. By scanning the position of the delay stage, the probe beam arrives at various time delays on the sample with respect to the pump beam, so that the temporal magnetization information of the sample can be measured. To measure the small polarization rotation in the probe induced by MOKE, a photo-elastic modulator (PEM) and lock-in detection is used. The PEM modulates the polarization of the probe beam at 50 kHz. After the reflection off the sample, the probe beam goes through an analyzer, converting polarization changes into intensity changes. The intensity of the probe beam is then measured with a Si photo-detector (Thorlabs PDB 450A-AC). By sending this intensity signal into a lock-in amplifier referenced at twice the PEM frequency (100 kHz), the polarization rotation caused by the magnetization can be obtained. All TR-MOKE scans are measured with an external magnetic field set in each of two opposite directions. Then the difference of the two scans is plotted, to cancel any non-magnetic contribution. At the optical wavelength of 810 nm, our measurement is only sensitive to the magnetization of the FeCo sublattice [28]. An external magnetic field of 200 Oe is applied during measurements in order to reset the magnetization between electrical pulses. Because the compensation temperature of GdFeCo is below room temperature, no transition across compensation occurs during optical or electrical heating. This means that the external magnetic field will always tend to align the magnetic moments back to the original direction, excluding the possibility of the external field assisting the switching [29].

**Optical absorption calculation.** To calculate the optical energy absorbed in the GdFeCo stack, we use a multilayer absorption calculation [14]. The electric field inside the stack is obtained through the transfer matrix method. The absorption was then obtained by calculating the divergence of the Poynting vector inside the stack. The resulting absorption per layer is reported in Table 1 in the Extended Data. The absorption profile is shown in Extended Data Figure 6. The whole GdFeCo stack absorbs 35% of the incident energy.

**Electrical pulse absorption calculation.** We determine the electrical energy absorbed in the GdFeCo section by two steps. First, we calculate the attenuation of the electrical pulse when propagating on the CPS to the GdFeCo section. We take the Fourier transform of the electrical pulse (voltage) $V(t)$ in time domain to get the frequency domain spectrum $\tilde{V}(\omega)$. The energy spectral density is then proportional to $|\tilde{V}(\omega)|^2$ (Extended Data Figure 5). The voltage on the CPS for an individual frequency $\omega$ signal at a given position $x_0$ away from the photo-switch is given by

$$\tilde{V}(\omega, x_0) = \exp\left(-\int_0^{x_0} \gamma dx\right) * \tilde{V}(\omega, 0)$$

where $\gamma$ is the propagation constant defined as follows

$$\gamma = \sqrt{(R + j\omega L)(G + j\omega C)}$$

*R, L, G, C* are the resistance, inductance, conductance and capacitance per unit length for the CPS. $R$ is estimated to be $10^4 \, \Omega/m$ and $10^3 \, \Omega/m$ for 5 μm wide and 50 μm wide CPS regions. $G$ is estimated to be 0.14 S/m and 0.014 S/m for 5 μm wide and 50 μm wide CPS regions. $L$ and $C$ can be calculated with the equations given in ref. 30. Due to non-perfect lithography, the impedance of the 5 μm wide and 50 μm wide

CPS are slightly different. We assume a gradual linear change of $\gamma$ across the taper region. The energy attenuation for a single frequency $\omega$ is then given by

$$\alpha_1(\omega) = \exp\left(-2Re\left(\int_0^{x_0} \gamma dx\right)\right) * Z_1(\omega)/Z_2(\omega)$$

where $Z_i$ is the frequency dependent impedance for 50 μm wide CPS ($Z_1$) and 5 μm wide CPS ($Z_2$), defined as

$$Z = \sqrt{(R + j\omega L)/(G + j\omega C)}$$

The frequency dependent attenuation before the GdFeCo load $\alpha_1(\omega)$ is plotted in Extended Data Figure 6.a.

As a second step, we calculate the absorption of the electrical pulse in the GdFeCo load. We use a multilayer absorption calculation, using the same method as in the optical absorption calculation. Here we assume that the electro-magnetic wave travels from the gold CPS into a thin layer (4 μm) of GdFeCo CPS, and then exits back into the gold CPS line. The effective complex refractive index is given by

$$n(\omega) = Conjugate\left(\frac{\gamma * c}{\omega * j}\right)$$

where c is the speed of light. The difference of complex refractive indices of the GdFeCo section and gold CPS section comes from the difference in $R$. For the GdFeCo CPS section, $R$ is estimated to be $2.48 * 10^7$ Ω/m instead of $10^4$ Ω/m for the 5 μm wide gold section. The absorption $\alpha_2(\omega)$, reflection and transmission across the GdFeCo load are plotted in Extended Data Figure 6.b.

Finally, the total absorption in the GdFeCo load can be calculated as follows.

$$\alpha = \frac{\int |\tilde{V}(\omega)|^2 * \alpha_1(\omega) * \alpha_2(\omega) d\omega}{\int |\tilde{V}(\omega)|^2 d\omega}$$

We calculate the total absorption in the GdFeCo to be 13%. The total energy carried by the initial electrical pulse is estimated by $\int I^2 * Z dt$ = 4.3 nJ, which means that we deliver about 570 pJ of electrical energy into the GdFeCo load.

**References**

1. Žutić, I., Fabian, J. & Das Sarma, S. Spintronics: Fundamentals and applications. *Reviews of Modern Physics* 76, 323-410 (2004).

2. Kirilyuk, A., Kimel, A. & Rasing, T. Ultrafast optical manipulation of magnetic order. *Reviews of Modern Physics* 82, 2731-2784 (2010).

3. Gerrits, T., van den Berg, H., Hohlfeld, J., Bär, L. & Rasing, T. Ultrafast precessional magnetization reversal by picosecond magnetic field pulse shaping. *Nature* 418, 509-512 (2002).

4. Garello, K. et al. Ultrafast magnetization switching by spin-orbit torques. *Appl. Phys. Lett.* 105, 212402 (2014).

5. Zhao, H. et al. Sub-200 ps spin transfer torque switching in in-plane magnetic tunnel junctions with interface perpendicular anisotropy. *Journal of Physics D: Applied Physics* 45, 025001 (2011).

6. Cubukcu, Murat. et al. Ultra-fast magnetization reversal of a three-terminal perpendicular magnetic tunnel junction by spin-orbit torque. *arXiv preprint arXiv:1509.02375* (2015).

7. Akerman, J. Toward a Universal Memory. *Science* 308, 508-510 (2005).

8. Behin-Aein, B., Datta, D., Salahuddin, S. & Datta, S. Proposal for an all-spin logic device with built-in memory. *Nature Nanotechnology* 5, 266-270 (2010).

9. Mistry, K. et al. A 45nm logic technology with high-k+ metal gate transistors, strained silicon, 9 Cu interconnect layers, 193nm dry patterning, and 100% Pb-free packaging. *IEEE International Electron Devices Meeting* (pp. 247-250). IEEE, 2007.

10. Beaurepaire, E., Merle, J., Daunois, A. & Bigot, J. Ultrafast Spin Dynamics in Ferromagnetic Nickel.*Phys. Rev. Lett.* 76, 4250-4253 (1996).

11. Stanciu, C. et al. All-Optical Magnetic Recording with Circularly Polarized Light. *Phys. Rev. Lett.*99, (2007).

12. Radu, I. et al. Transient ferromagnetic-like state mediating ultrafast reversal of antiferromagnetically coupled spins. *Nature* 472, 205-208 (2011).

13. Ostler, T. et al. Ultrafast heating as a sufficient stimulus for magnetization reversal in a ferrimagnet.*Nature Communications* 3, 666 (2012).

14. Gorchon, Jon. et al. The role of electron temperature in the helicity-independent all-optical switching of GdFeCo. *arXiv preprint arXiv:1605.09764* (2016).

15. Ketchen, M. et al. Generation of subpicosecond electrical pulses on coplanar transmission lines. *Appl. Phys. Lett.* 48, 751 (1986).

16. Gupta, S. et al. Subpicosecond carrier lifetime in GaAs grown by molecular beam epitaxy at low temperatures. *Appl. Phys. Lett.* 59, 3276 (1991).

17. Wilson, R. B.. et al. Electric Current Induced Ultrafast Demagnetization. *arXiv preprint arXiv:1609.00758* (2016).

18. Koopmans, B. et al. Explaining the paradoxical diversity of ultrafast laser-induced demagnetization. *Nature materials* 9, 259-265 (2010).

19. Tas, G. & Maris, H. Electron diffusion in metals studied by picosecond ultrasonics. *Phys. Rev. B* 49, 15046-15054 (1994).

20. Wilson, R. B.. et al. Ultrafast Magnetic Switching of GdFeCo with Electronic Heat Currents. *arXiv preprint arXiv:1609.05155* (2016).

21. Chen, J. Y., He, L., Wang, J. P., & Li, M. Picosecond all-optical switching of magnetic tunnel junctions. *arXiv preprint arXiv:1607.04615*. (2016)

22. 2013 ITRS. *International Technology Roadmap for Semiconductors* (2016). at <http://www.itrs2.net/2013-itrs.html>

23. Liu, H. et al. Ultrafast switching in magnetic tunnel junction based orthogonal spin transfer devices. *Appl. Phys. Lett.* 97, 242510 (2010).

24. Zeng, Z. et al. Effect of resistance-area product on spin-transfer switching in MgO-based magnetic tunnel junction memory cells. *Appl. Phys. Lett.* 98, 072512 (2011).

25. Zhao, H. et al. Sub-200 ps spin transfer torque switching in in-plane magnetic tunnel junctions with interface perpendicular anisotropy. *Journal of Physics D: Applied Physics* 45, 025001 (2011).

26. Garello, K. et al. Ultrafast magnetization switching by spin-orbit torques. *Appl. Phys. Lett.* 105, 212402 (2014).

27. Wong, H.-S. P. et al. *Stanford Memory Trends*, https://nano.stanford.edu/stanford-memory-trends, accessed September 2, 2016.

28. Hohlfeld, J. et al. Fast magnetization reversal of GdFeCo induced by femtosecond laser pulses. *Phys. Rev. B* 65, (2001).

29. Stanciu, C. et al. Subpicosecond Magnetization Reversal across Ferrimagnetic Compensation Points. *Phys. Rev. Lett.* 99, (2007).

30. Erli Chen, & Chou, S. Characteristics of coplanar transmission lines on multilayer substrates: modeling and experiments. *IEEE Transactions on Microwave Theory and Techniques* 45, 939-945 (1997).

**Acknowledgements**

This work was supported by the National Science Foundation Center for Energy Efficient Electronics (sample fabrication and laser technology) and by the Director, Office of Science, Office of Basic Energy Sciences, Materials Sciences and Engineering Division, of the U.S. Department of Energy under Contract No. DE-AC02-05-CH11231 within the Nonequilibrium Magnetic Materials Program [MSMAG] (experimental operations). We also acknowledge partial support in the early stages of this project for conceptual design and photoconducting switch development by C-SPIN: one of the six SRC STARnet Centers, sponsored by MARCO and DARPA.

**Author Contributions**

Y.Y., R. B. W., J. G., designed the experiments under the supervision of J.B..  Y. Y. fabricated the devices with help from J. G. and C. L.. Y.Y., J.G. and R.B.W., performed the optical and electrical magnetization switching experiments. C.L. sputtered and characterized the magnetic films under the supervision of S.S.. Y.Y. analyzed the experimental data with help from R.B.W. and J.G.. Y.Y. wrote the manuscript with input from all authors.

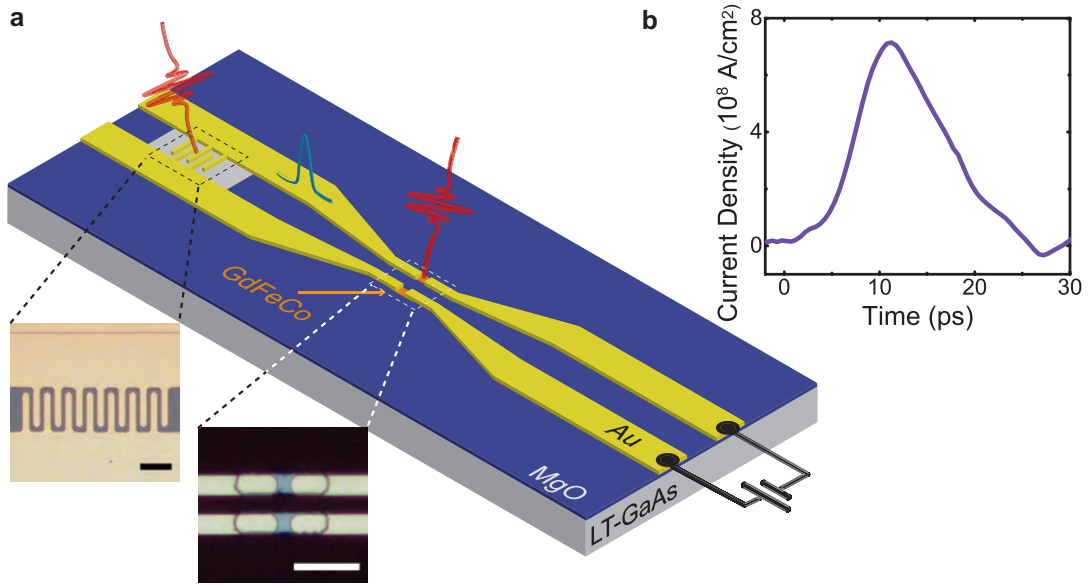

**Figure 1. Schematic of the CPS device and characterization of electrical pulse.**

**a.** Schematic of electrical switching experiment. The photo-switch is illuminated with laser pulses while biased with a DC source. Magnetization dynamics of GdFeCo is monitored with TR-MOKE. **Left Inset:** Optical image of the photo-switch. During laser illumination, photo-excited carriers in low temperature GaAs conduct current across the gap, generating a transient electrical pulse propagating both directions. **Right Inset:** Optical image of GdFeCo section of CPS. Scale bar: 20 μm. **b.** Calculated temporal current density profile through the GdFeCo section, based on the temporal current profile measured with Protemics Spike probe positioned 1 mm before the GdFeCo section.

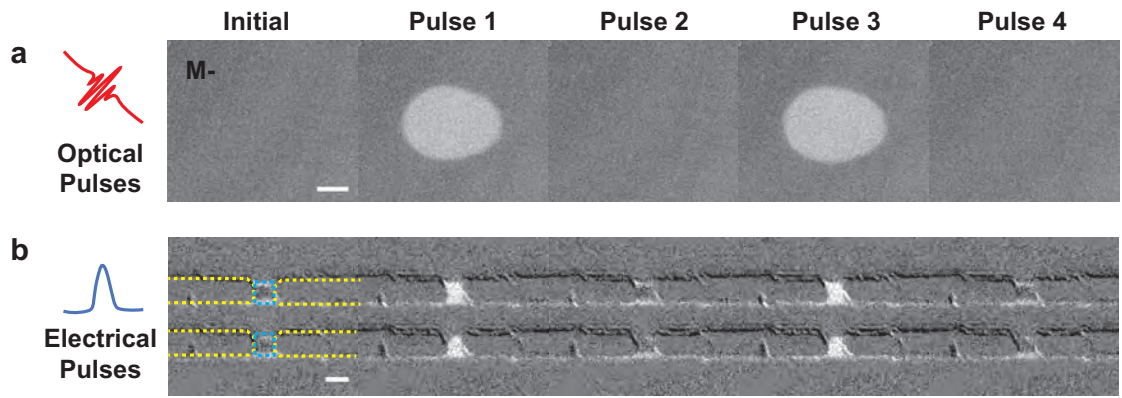

**Figure 2. Single-shot optical and electrical switching of GdFeCo.**

**a.** Differential MOKE images of bare GdFeCo film after sequential 6.4 ps optical pulse irradiation. Absorbed fluence is 1.8 mJ/cm². After each optical pulse, the magnetization of GdFeCo toggles to the opposite direction. The contrast indicates change in magnetization. **b.** Differential MOKE images of GdFeCo CPS section after sequential 9 ps electrical pulse excitation. After each electrical pulse, the magnetization of GdFeCo toggles. Yellow and blue dash lines indicate gold CPS and GdFeCo section. Scale bar: 5 μm.

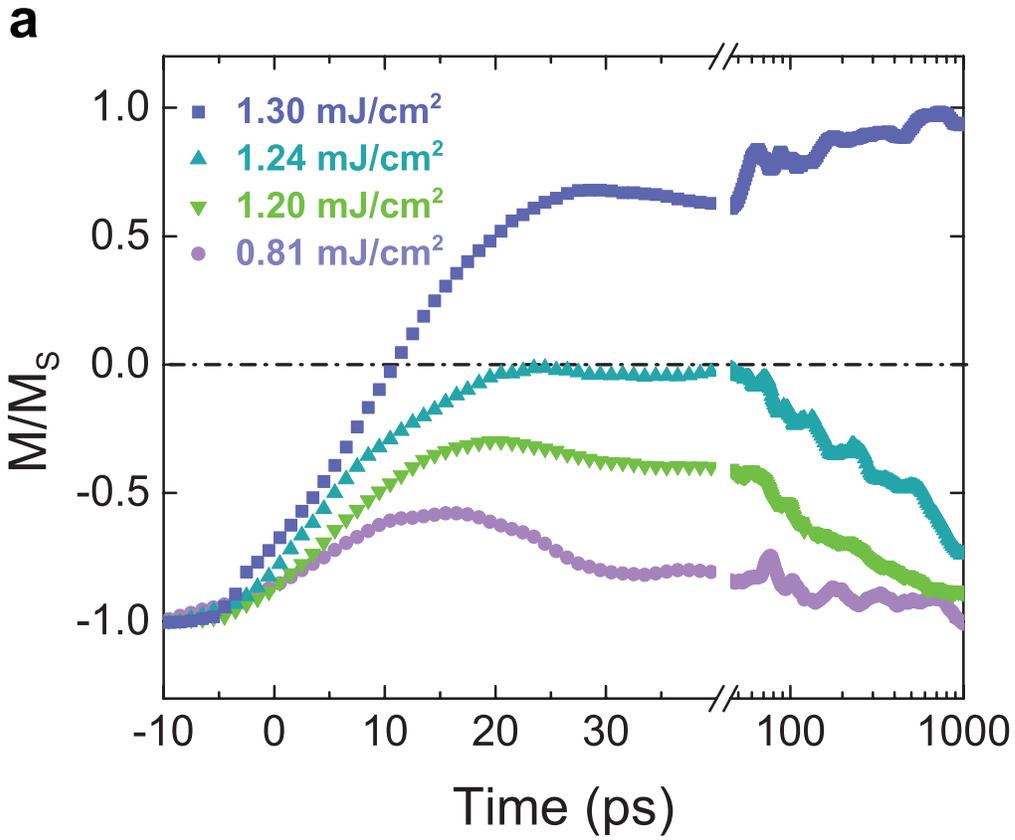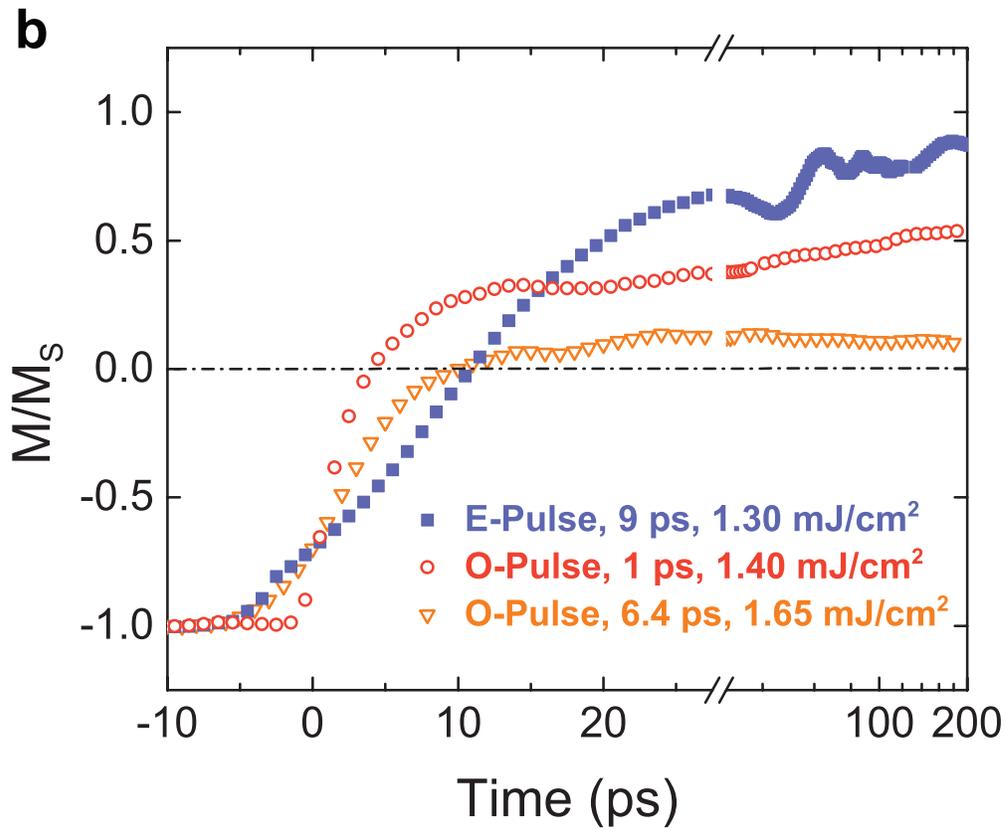

**Figure 3. Magnetization dynamics of GdFeCo after electrical and optical excitations.**

**a.** Electrical demagnetizing and switching of GdFeCo. All fluences are calculated absorbed fluences. With increasing electrical pulse amplitude, the GdFeCo demagnetization amplitude increases, and eventually switches around 10 ps.

**b.** Comparison of electrical and optical switching of GdFeCo. Both electrical and optical fluences are calculated absorbed fluences.

**Extended Data**

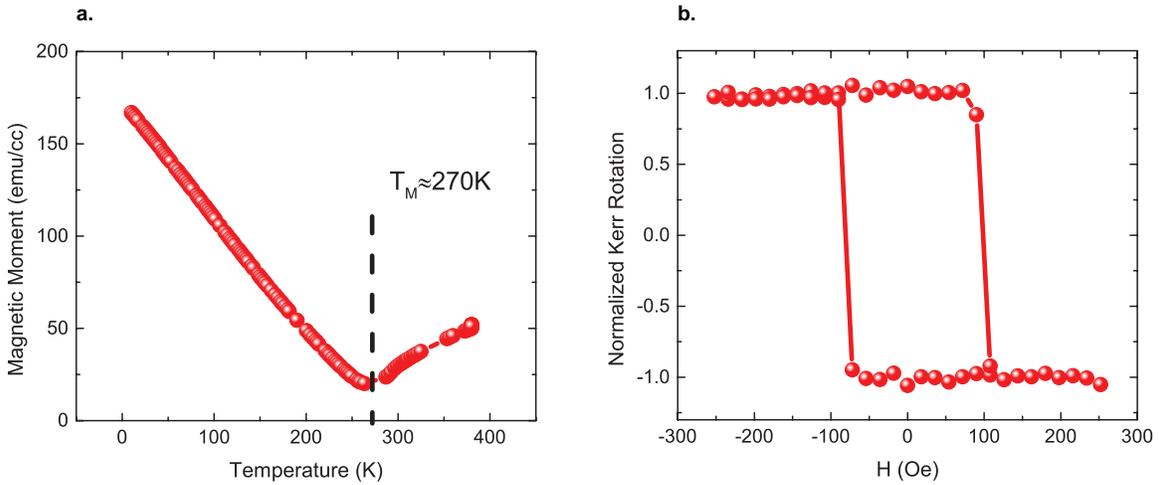

**Extended Data Figure 1. Magnetic Properties of the GdFeCo stack. a.** Magnetic moment of the GdFeCo film measured with a superconducting quantum interference device (SQUID) at various temperatures. An out-of-plane external magnetic field of 500 Oe is applied during all the measurements. **b.** Magnetic hysteresis curve of the GdFeCo film at room temperature, measured with the MOKE microscope.

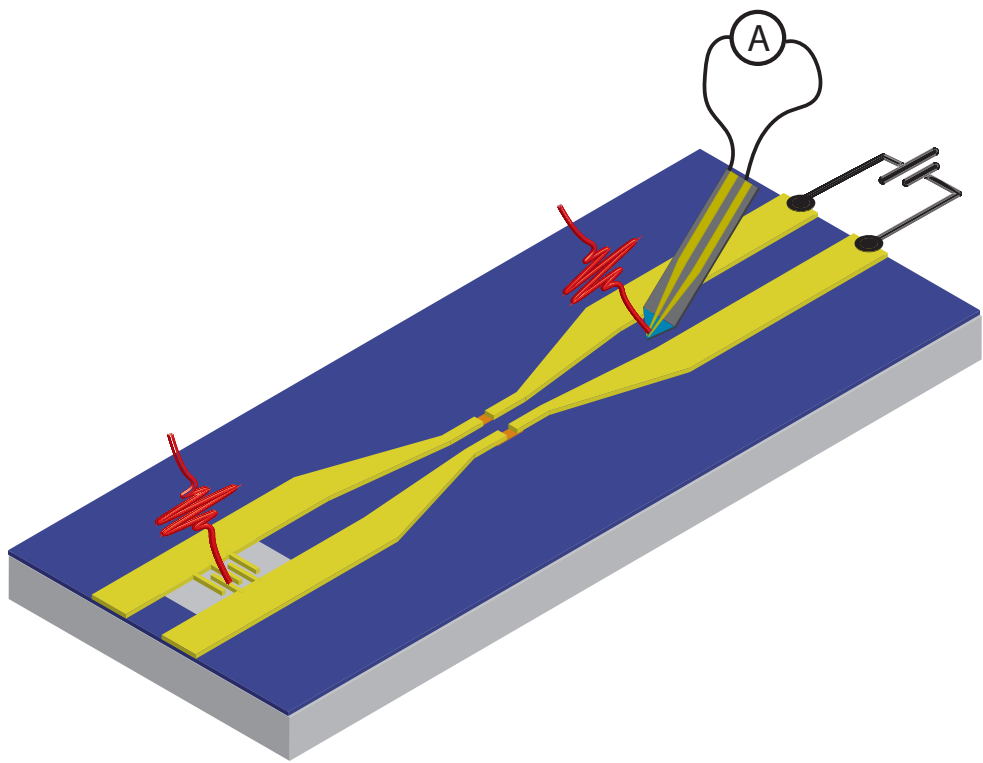

**Extended Data Figure 2. Schematic of the experimental setup for measuring the temporal profile of the electrical pulse.** A pump beam is on the photoconductive switch to generate the pulse. A probe beam is on the photoconductive switch at the tip of Protemics Spike probe. A current amplifier is connected to Protemics Spike probe to measure the average current induced by the transient electric field at the tip of the probe. Current profile in Extended Figure 3 is measured before the GdFeCo section as oppose to this schematic.

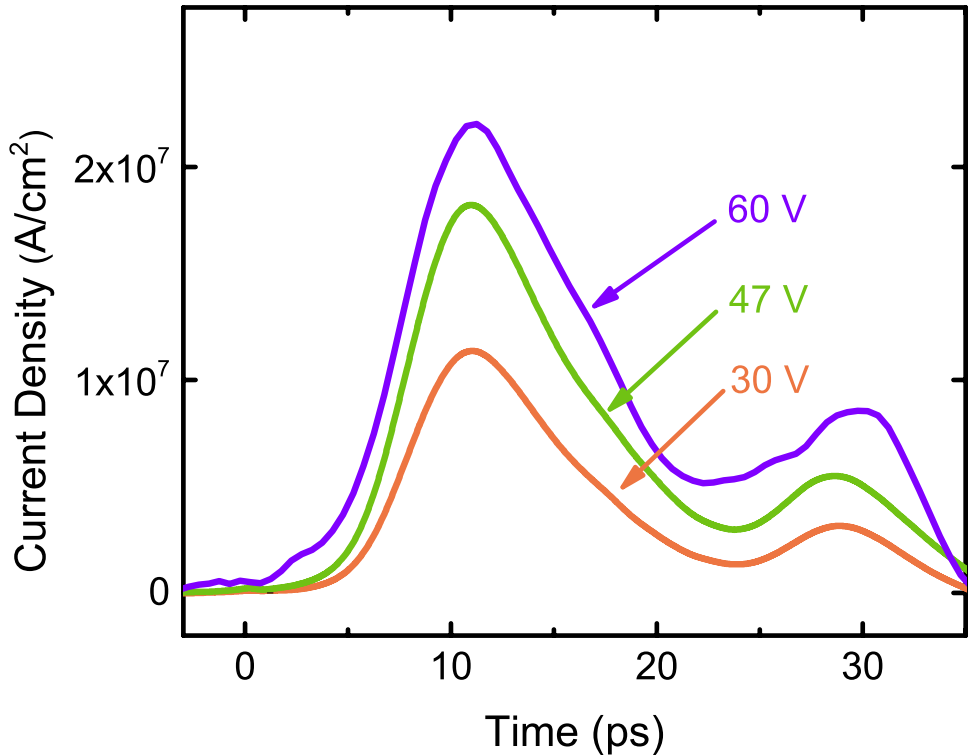

**Extended Data Figure 3. Temporal current density profiles in CPS**. Temporal current density profile generated by the photo-switch in 50 μm wide CPS Au line with different biases, as measured with Protemics Spike probe positioned 1mm before the GdFeCo section. The smaller electrical pulse following the main peak is attributed to electrical reflection from the GdFeCo section in the CPS.

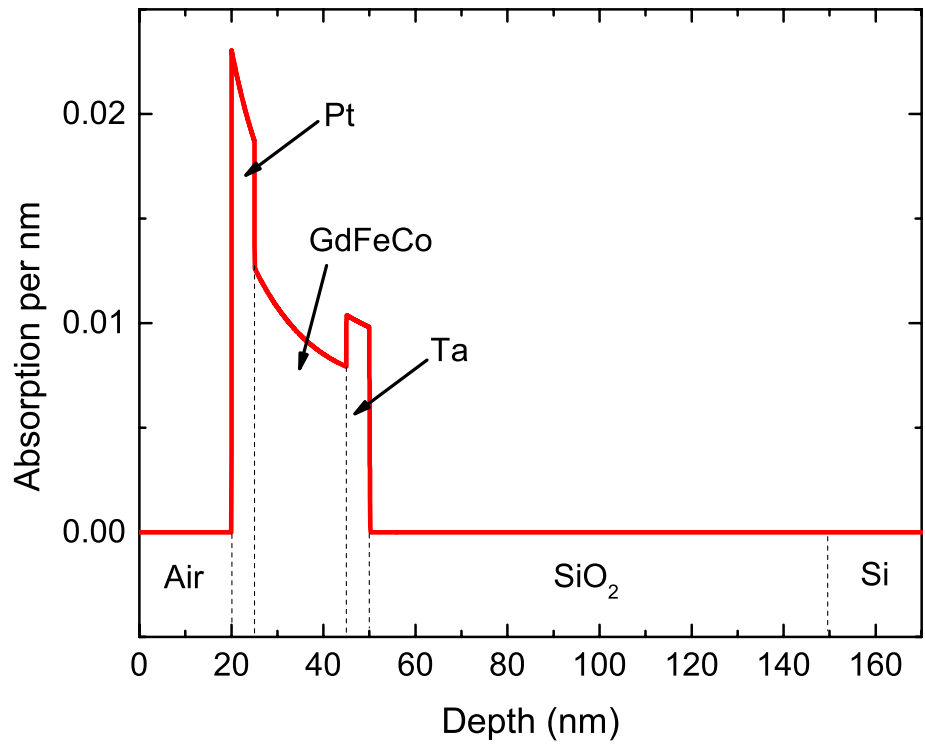

**Extended Data Figure 4. Calculated optical absorption profile for the Pt/GdFeCo/Ta stack at 810 nm wavelength.** Incident angle is 40 degrees to normal.

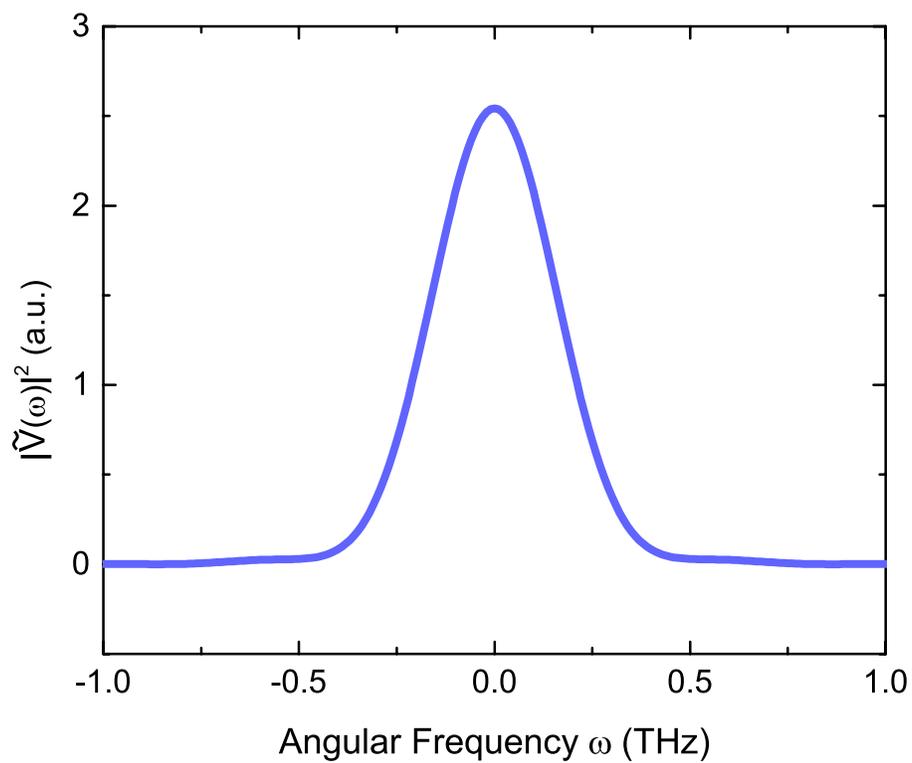

**Extended Data Figure 5. Energy spectral density of the electrical pulse**, calculated as the square of the Fourier transform of the electrical pulse in the time-domain.

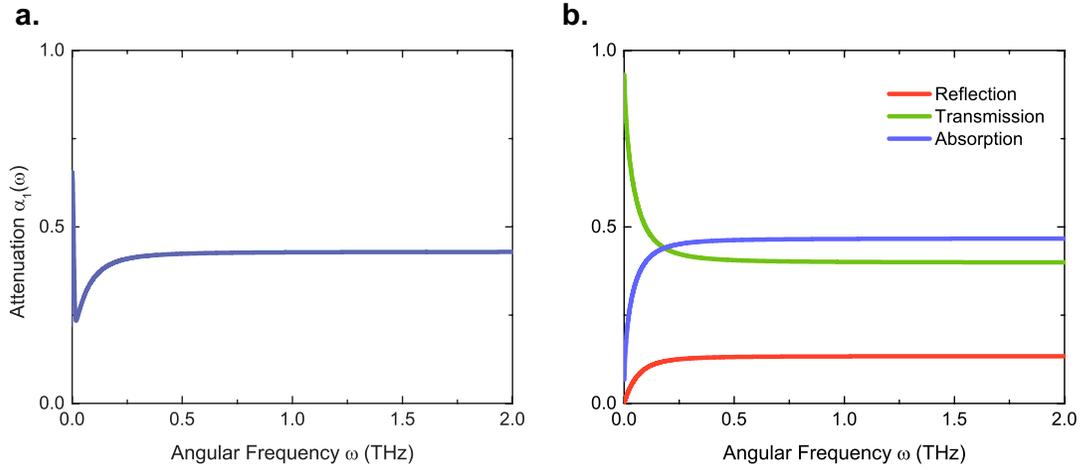

**Extended Data Figure 6. Frequency dependent absorption calculation on CPS**

**a.** Attenuation of different frequency components on the CPS before the GdFeCo section. **b.** Reflection, transmission and absorption across the GdFeCo section at various frequencies.

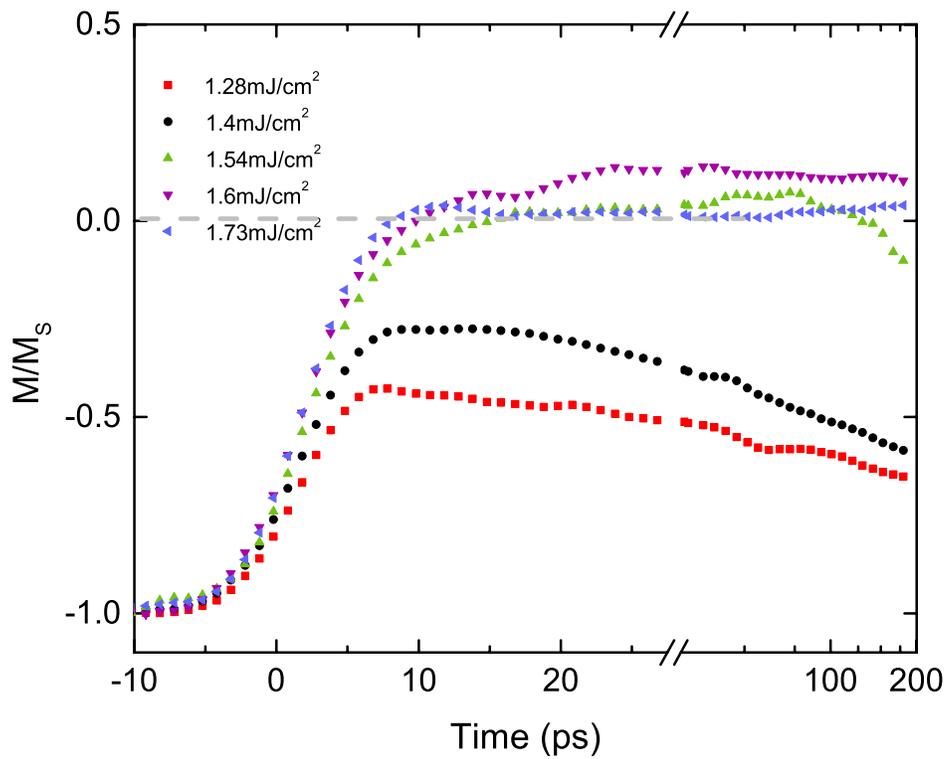

**Extended Data Figure 7. TR-MOKE for 6.4 ps optical pulses with different fluences.** The critical fluence for switching is around 1.6 mJ/cm². The GdFeCo film becomes completely demagnetized above 1.73 mJ/cm² because the lattice temperature exceeds Curie temperature.

**Extended Data Table 1. Optical multilayer calculation parameters and results**

| Material | Thickness (nm) | Refractive Index | Absorption (%) |
|---|---|---|---|
| Air | - | 1 | - |
| Pt | 5 | 2.85+4.96i | 10.4 |
| GdFeCo | 20 | 2.66+3.6i | 19.4 |
| Ta | 5 | 3.43+3.66i | 5 |
| SiO$_2$ | 100 | 1.45 | 0 |
| Si | - | 3.696+0.0047i | - |